\documentclass[aip,amsmath,amssymb,reprint]{revtex4-2}

\usepackage{graphicx}
\usepackage{float}
\usepackage{multirow}
\usepackage{natbib,twoopt}
\usepackage[breaklinks=true]{hyperref}
\usepackage{xcolor}
\usepackage{amssymb}
\usepackage{stackengine,graphicx}
\usepackage{amsmath}
\usepackage{hyperref}
\usepackage{color}
\usepackage{stackengine}
\usepackage{subfigure}
\usepackage{verbatim}
\usepackage{mathtools}
\usepackage{braket}
\usepackage{bm}

\newcommand{\rot}{\mathop{\rm rot}\nolimits}
\newcommand{\divv}{\mathop{\rm div}\nolimits}
\newcommand{\df}[2]{\frac{\partial #1}{\partial #2}}

\newcommand{\eps}{\varepsilon}

\newcommand{\vc}[1]{{\bf #1}}
\newcommand{\skob}[1]{\left( #1\right)}

\renewcommand{\c}{\chi}

\usepackage{hyperref}
\usepackage{float}



\begin{document}

\title{Unbounded Tellegen Response in Media with Multiple Resonances}

\author{Timur Z. Seidov}
\email{timur.seidov@metalab.ifmo.ru}
\affiliation{School of Physics and Engineering, ITMO University, Saint  Petersburg, Russia}

\author{Maxim A. Gorlach}
\affiliation{School of Physics and Engineering, ITMO University, Saint Petersburg, Russia}

\begin{abstract}
Tellegen response is a special type of nonreciprocal magneto-electric coupling which long remained elusive in photonics and extremely weak in condensed matter. It is widely accepted that the Tellegen coefficient is restricted by $\chi^2<\epsilon\mu$, where $\eps$ and $\mu$ are permittivity and permeability of the material. Here, we demonstrate that this restriction is lifted in the medium with several close resonances, which provides a theoretical foundation for giant Tellegen response.
\end{abstract}

\maketitle

\section{Introduction}


Artificially structured media enable exotic optical phenomena by tailoring light-matter interactions at subwavelength scales~\cite{Eleftheriades2005,Marques2008,Capolino2009}. An example of such kind is bianisotropy also known as magneto-electric coupling~\cite{Serdyukov_2001}. In the simplest isotropic case it is captured by the constitutive relations 
\begin{gather}
    \vc D = \eps \vc E + (\c + i \kappa) \vc H \:, \label{eq:Const1}\\
    \vc B = (\c - i \kappa) \vc E +\mu \vc H\:,\label{eq:Const2}
\end{gather}
where $\eps$ and $\mu$ are permittivity and permeability of the medium, while $\kappa$ and $\chi$ capture the effect of bianisotropy and are referred to as chirality and Tellegen response. In the absence of losses, both parameters are real.


Chirality requires breaking of inversion symmetry and is ubiquitous in nature arising in many organic molecules with spiral structure. In contrast, Tellegen response is more exotic and occurs only if both inversion and time reversal symmetry of the medium are broken. Here, we focus on the latter response which was initially postulated by Tellegen in 1948~\cite{tellegen1948gyrator} and has since been a subject of active investigation. 

Initially, the very existence of the Tellegen response was disputed~\cite{Lakhtakia_Weiglhofer,kamenetskii2008tellegenparticlesreallyexist,Lakhtakia2009}. However, this debate is now resolved, as there are multiple examples of Tellegen media in condensed matter including magnetoelectrics (e.g. $\mathrm{Cr}_2\mathrm{O}_3$ \cite{krichevtsov1993spontaneous}), multiferroics and topological insulators~\cite{Nenno2020,Sekine2021}. Electromagnetics of those materials mirrors the equations of axion electrodynamics~\cite{Wilczek1987,Nenno2020}. However, the typical values of $\chi$ are quite small ranging from $10^{-6}$ to $10^{-3}$~\cite{fiebig2005revival}.



It is a commonly accepted notion \cite{PhysRev.168.574,Shuvaev2011,PhysRevA.84.063822,lindell1995methods,Lindell_1994,Serdyukov_2001} that the Tellegen response of isotropic medium is constrained by 
\begin{equation}\label{eq:TellegenRestriction}
 |\c|\leq\sqrt{\eps\mu}
\end{equation}
Due to the small values of $\chi$ in condensed matter, this constraint has never been approached. However, the situation has changed with the recent theoretical~\cite{Silv2023,Shaposhnikov2023,SafaeiJazi2024} and experimental~\cite{yang2024tellegenresponsesmetamaterials} studies in photonics which suggested extremely strong Tellegen response of the order of $0.1-1$. Although experimentally reported system is not yet bulk metamaterial, this creates tension between the established bound on the Tellegen coefficient and recent experimental findings.

In this article, we prove that the restriction Eq.~\eqref{eq:TellegenRestriction} on the Tellegen coefficient can be lifted if the medium possesses multiple spectrally close resonances, which lays the theoretical foundation for achieving strong Tellegen response.

First, we reiterate the logic of the constraint Eq.~\eqref{eq:TellegenRestriction}. In the absence of sources and in the CGS system of units Maxwell's equations take the form
\begin{gather}
    \rot \vc H =\frac{1}{c}\df{\vc D}{t}\:, \\
    \rot \vc E =-\frac{1}{c}\df{\vc B}{t}\:, \\
    \divv \vc D=0\:, \\
    \divv \vc B=0
\end{gather}
with the constitutive relations Eqs.~\eqref{eq:Const1},\eqref{eq:Const2}. In monochromatic case, the latter two equations are a consequence of the first two and hence can be omitted. Without loss of generality, we denote the direction of propagation as $z$-axis and introduce the circular basis $\vc e_{\pm}=\vc e_{x}\pm i\vc e_{y}$. The resulting equations for the amplitudes $E_{\pm}$, $H_{\pm}$ of the circularly polarized modes read
\begin{gather}
    \eps E_\pm + [\c\mp i n] H_\pm=0 \:, \\
    [\c\pm i n] E_\pm + \mu H_\pm=0 \:,
\end{gather}
$n\equiv c k/\omega$ being a refractive index. Solving these equations yields the refractive index $n=\sqrt{\eps\mu-\c^2}$. Once $|\c| > \sqrt{\eps\mu}$, the refractive index becomes purely imaginary suppressing wave propagation. However, this does not mean that $|\chi|>\sqrt{\eps\mu}$ is impossible. It only means that the medium does not sustain the propagating modes. Such situation happens, for instance, in plasma below the plasma frequency and does not lead to any inconsistencies.

Therefore, to derive the constraint on $\chi$, we compute the field energy in Tellegen medium neglecting losses and frequency dispersion and presenting energy density as a quadratic form~\cite{lindell1995methods,Lindell_1994,Serdyukov_2001}
%
\begin{equation}
    W=\frac{1}{16\pi}
    \begin{pmatrix}
        \vc E & \vc H
    \end{pmatrix}
    \begin{pmatrix}
        \eps & \c \\ \c & \mu
    \end{pmatrix}
    \begin{pmatrix}
        \vc E \\ \vc H
    \end{pmatrix}^* \label{eq:energy_density}
\end{equation}

As the field energy must be non-negative, the material parameters should satisfy
\begin{equation}
    \c^2\leq \eps\mu \:, \quad \eps>0\:, \quad \mu>0, \label{eq:ineq_lossless}
\end{equation}
yielding the restriction Eq.~\eqref{eq:TellegenRestriction}.

In the same spirit, we may consider a single Tellegen meta-atom~\cite{SafaeiJazi2024} acquiring both electric $\vc d$ and magnetic $\vc m$ dipole moments in an external field. The dipole moments of such particle are expressed via polarizabilities taken scalar for simplicity:
\begin{gather}
    \vc d = \alpha^{ee} \vc E + \alpha^{em} \vc H\:, \\
    \vc m = \alpha^{me} \vc E + \alpha^{mm} \vc H\:.
\end{gather}
In case of Tellegen response both magneto-electric polarizabilities are real:  $\alpha^{em}=\alpha^{me}\in \rm I\!R$. Neglecting frequency dispersion, we present the potential energy in the external field in the form
\begin{equation}
    U=-\frac{1}{2}\left(\alpha^{ee} \vc E^2 + \alpha^{mm} \vc H^2 \right)-\alpha^{em}\vc E\cdot \vc H\:.
\end{equation}
As the potential energy should be non-positive for the arbitrary configuration of external field,  $\alpha^{ee}\geq 0$, $\alpha^{mm}\geq 0$, and
\begin{equation}\label{eq:PolarizabilityRest}
  |\alpha^{em}|\leq\sqrt{\alpha^{ee}\alpha^{mm}}\:,  
\end{equation}
mirroring the constraints Eq.~\eqref{eq:ineq_lossless}.

Hence, in the absence of dispersion, i.e. away from the resonances of the medium, both bulk Tellegen response and magneto-electric polarizability of a single Tellegen particle are constrained in agreement with the early works.


We now advance this description by introducing a single resonance in polarizabilitity. It is associated with the meta-atom eigenmode characterized by non-vanishing, collinear and in-phase electric and magnetic dipole moments. In such case, the analytical description of the meta-atom response should be based on its quasi-normal modes~\cite{Kuipers2017,Lalanne2018}. However, to avoid the difficulties associated with the leaky nature of quasi-normal modes, we instead adopt conceptually similar quantum-mechanical methodology for computing electromagnetic response in the first-order of perturbation theory neglecting losses~\cite{Barron_2004}.

Technically, we analyze time-dependent Schr{\"o}dinger equation
\begin{equation}
    \left(i\hbar \frac{\partial}{\partial t} - \hat H \right) \ket{\psi}=\hat V  \ket{\psi} \:, \label{eq:Schrodinger}
\end{equation}
where $\hat H$ is an unperturbed Hamiltonian of the system, while $\hat V$ quantifies the interaction with the incident electromagnetic field. We suppose that the system is electrically small and employ dipole approximation, assuming that the matrix elements of electric $\vc d$ and magnetic $\vc m$ dipole moments are significant, while the contribution of higher-order multipoles is negligible, therefore allowing us to recast the interaction Hamiltonian in the following form
\begin{multline}
    2\hat V= -\hat d_\beta \hat E_{0\beta} e^{-i\omega t } - \hat d_\beta \hat E_{0\beta}^* e^{i\omega t }  \\  - \hat m_\beta \hat B_{0\beta} e^{-i\omega t } - \hat m_\beta \hat B_{0\beta}^* e^{i\omega t }\:. \label{eq:interaction}
\end{multline}
Here $\vc E_0$ and $\vc B_0$ are complex amplitudes of the incident field, Greek indices denote vector components and the summation is performed over repeated indices. General first-order perturbation theory for the polarizabilities is provided in the appendix. In the simplest case of isotropic two-level system the set of polarizabilities reads
\begin{gather}
    \alpha^{ee}=\frac{2\omega_{10}}{3\hbar}\frac{||\vc d_{10}||^2}{\omega_{10}^2-\omega^2}\:, \\
    \alpha^{mm}=\frac{2\omega_{10}}{3\hbar}\frac{||\vc m_{10}||^2}{\omega_{10}^2-\omega^2}\:, \\
    \Im(\alpha^{em})=\frac{2\omega}{3\hbar}\frac{\Im{(\vc d_{10},\vc m_{10})}}{\omega_{10}^2-\omega^2}\:, \\
    \Re(\alpha^{em})=\frac{2\omega_{10}}{3\hbar}\frac{\Re{(\vc d_{10},\vc m_{10})}}{\omega_{10}^2-\omega^2}\:.
\end{gather}
%
Here, $(\vc a ,\vc b)$ is a Hermitian product of the complex vectors $\vc a$ and $\vc b$, i.e. $\vc a^*\cdot \vc b$, $||\vc a||^2\equiv(\vc a,\vc a)$,  $\omega_{jn}$ is the difference between $j$-th and $n$-th eigenfrequencies of the unperturbed system, $\alpha^{me}=\left(\alpha^{em}\right)^*$ and $\Re$, $\Im$ symbols stand for real and imaginary parts of the complex number. 

Real and imaginary parts of magneto-electric polarizability are readily identified as Tellegen response and chirality. From these expressions, we conclude that
\begin{multline}
    |\Re(\alpha^{em})|=\frac{2\omega_{10}}{3 \hbar} \frac{|\Re{(\vc d_{10},\vc m_{10})}|}{|\omega_{10}^2-\omega^2|}\leq \\ \frac{2\omega_{10}}{3 \hbar} \frac{||\vc d_{10}|| \: ||\vc m_{10}||}{|\omega_{10}^2-\omega^2|}=\sqrt{\alpha^{ee}\alpha^{mm}}\:,
\end{multline}
and the equality holds when $\vc d_{10}$ and $\vc m_{10}$ are in phase and parallel.

Thus, in the presence of a single resonance, we recover the same restriction on magneto-electric polarizability Eq.~\eqref{eq:PolarizabilityRest}, even though electric and magnetic polarizabilities can become negative at high frequencies.

Essentially, this logic motivated the community to believe that the effective Tellegen response is restricted by Eq.~\eqref{eq:TellegenRestriction}, while Tellegen polarizability of a single particle is constrained by Eq.~\eqref{eq:PolarizabilityRest}. However, this treatment is incomplete as it ignores an essential physical aspect: any Mie-resonant dielectric particle as well as any material has more than a single resonance. As we prove below, the interplay of close resonances allows to overcome the restriction Eq.~\eqref{eq:PolarizabilityRest} paving a way to strong Tellegen nonreciprocity.




To illustrate that, we examine a system with two resonances (i.e. three-level system) using the same  approach detailed in the appendix. Assuming isotropic ground state, we recover the polarizabilities
\begin{gather}
    \alpha^{em}=\frac{2}{3\hbar}\Big[\frac{1}{\omega_{10}^2-\omega^2}(\omega_{10}\Re{(\vc d_{10},\vc m_{10})}+i\omega \Im{(\vc d_{10},\vc m_{10})})+ \nonumber \\ \frac{1}{\omega_{20}^2-\omega^2}(\omega_{20}\Re{(\vc d_{20},\vc m_{20})}+i\omega \Im{(\vc d_{20},\vc m_{20})})\Big],  \\
    \alpha^{ee}=\frac{2}{3\hbar}\Big[\frac{\omega_{10}}{\omega_{10}^2-\omega^2}||\vc d_{10}||^2 +\frac{\omega_{20}}{\omega_{20}^2-\omega^2}||\vc d_{20}||^2\Big], \\
    \alpha^{mm}=\frac{2}{3\hbar}\Big[\frac{\omega_{10}}{\omega_{10}^2-\omega^2}||\vc m_{10}||^2 +\frac{\omega_{20}}{\omega_{20}^2-\omega^2}||\vc m_{20}||^2\Big].
\end{gather}
%
We introduce the following notation for the resonant Lorentz factors
\begin{equation}
    P(\omega_{i0})=\frac{2}{3\hbar}\frac{\omega_{i0}}{\omega_{i0}^2-\omega^2}.
\end{equation}

For simplicity, we assume that the chiral response is absent, i.e. $\Im (\vc d_{i0},\vc m_{i0})=0$. Hence, $\vc d_{i0}$ and $\vc m_{i0}$ are in phase and without loss of generality can be assumed real vectors.
To check whether $|\Re(\alpha^{em})|\leq\sqrt{\alpha^{ee}\alpha^{mm}}$ is satisfied, we examine the sign of the difference
\begin{equation}
 \Delta=\Re^2(\alpha^{em})-\alpha^{ee}\alpha^{mm}\:,  
\end{equation}
where $\Delta>0$ would indicate violation of restriction Eq.~\eqref{eq:PolarizabilityRest}. Expanding expressions for polarizabilities yields
\begin{multline}
    \alpha^{ee}\alpha^{mm}= \\     P^2(\omega_{10})||\vc{d_{10}}||^2||\vc{m_{10}}||^2 + P^2(\omega_{20})||\vc{d_{20}}||^2||\vc{m_{20}}||^2 + \\
    P(\omega_{10})P(\omega_{20})(||\vc{d_{10}}||^2||\vc{m_{20}}||^2+||\vc{d_{20}}||^2||\vc{m_{10}}||^2)\:,
\end{multline}
\begin{multline}
    \Re^2(\alpha^{em})= \\
    P^2(\omega_{10})||\vc{d_{10}}||^2||\vc{m_{10}}||^2 \cos^2\theta_1 + \\ P^2(\omega_{20})||\vc{ d_{20}}||^2||\vc{m_{20}}||^2 \cos^2\theta_2 + \\
    2 P(\omega_{10})P(\omega_{20}) ||\vc{ d_{10}}||\:||\vc{m_{10}}||\:||\vc{ d_{20}}||\:||\vc{m_{20}}|| \cos \theta_1 \cos \theta_2,
\end{multline}
where $\theta_i$ is the angle between $\vc d_{i0}$ and $\vc m_{i0}$. For clarity, we assume that $\vc d_{i0}$ and $\vc m_{i0}$ are parallel, i.e. $\cos \theta_i=1$. In such scenario 
\begin{multline}
    \Delta=\Re^2(\alpha^{em})-\alpha^{ee}\alpha^{mm} = \\
    -P(\omega_{10})P(\omega_{20}) \: (||\vc{ d_{20}}||\:||\vc{m_{10}}||-||\vc{ d_{10}}||\:||\vc{m_{20}}||)^2.
\end{multline}
Expression in the brackets is always positive. Thus, the sign of the overall expression is defined by the sign of the two Lorentz factors
\begin{equation}
    -\frac{\omega_{10} \omega_{20}}{(\omega_{10}^2-\omega^2)(\omega_{20}^2-\omega^2)}.
\end{equation}

\begin{figure}
    \centering
    \includegraphics[width=1 \linewidth]{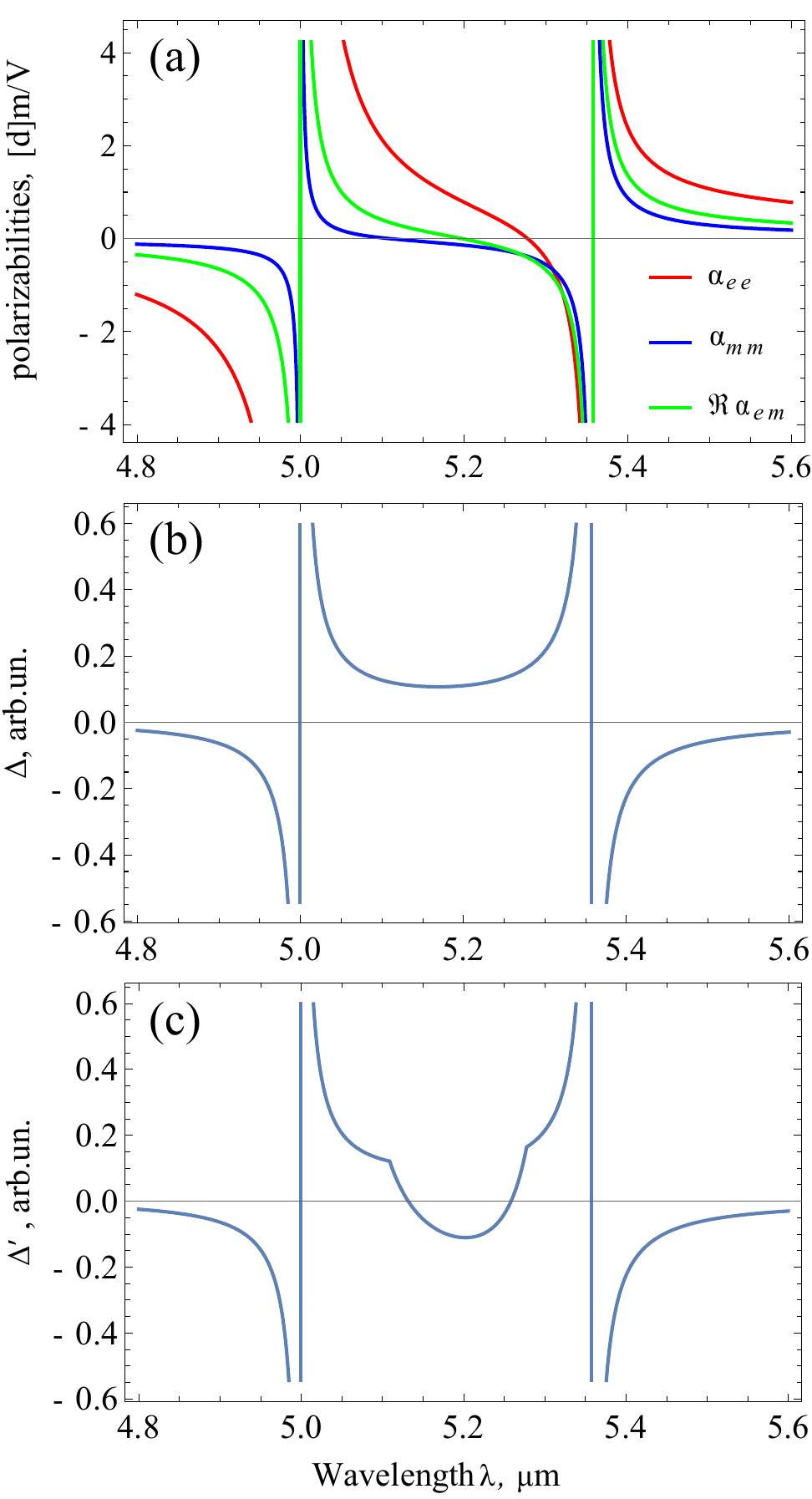}
    \caption{Probing the restrictions on the Tellegen polarizability of a meta-atom. The resonances correspond to those in the Weyl-based meta-atom~\cite{SafaeiJazi2024}, with wavelengths $\lambda_1=5$ $\mathrm{\mu}$m, $\lambda_2=5.36$ $\mathrm{\mu}$m. Matrix elements of the dipole moments are chosen $|d_{10}|=1.24, |d_{20}|=2.36, |m_{10}|=0.86, |m_{20}|=0.58$ in arbitrary units. (a) Calculated dependence of polarizabilities on wavelength. (b) Wavelength dependence of the difference $\Delta$. Violation of restriction corresponds to $\Delta>0$. (c) Wavelength dependence of the difference $\Delta'$, violation of restriction corresponds to $\Delta'>0$.}
    \label{fig:enter-label}
\end{figure}

If $\omega_{10}<\omega_{20}$, the difference $\Delta$ is negative for $\omega<\omega_{10}$ and $\omega>\omega_{20}$. However, in the region between the resonances $\omega_{10}<\omega<\omega_{20}$ the difference $\Delta$ becomes positive, and thus the restriction on the Tellegen polarizability Eq.~\eqref{eq:PolarizabilityRest} is lifted.

This conceptual conclusion is illustrated in Fig.~\ref{fig:enter-label}, which depicts the polarizabilities computed for parameters close to  Ref.~\cite{SafaeiJazi2024} [Fig.~\ref{fig:enter-label}(a)]. Figure~\ref{fig:enter-label}(b) shows that the difference $\Delta$ indeed takes positive values between the two resonances thus violating the celebrated restriction on the Tellegen response. Note that the above restriction is violated not due to the negative values of electric and magnetic polarizability, but rather due to the high $\alpha^{em}$, which is readily confirmed by plotting the difference $\Delta'=\Re^2(\alpha^{em})-|\alpha^{ee}\alpha^{mm}|$  in Fig.~\ref{fig:enter-label}(c).



To summarize, we have demonstrated that the Tellegen response of a non-reciprocal magneto-electric particle can exceed the bound Eq.~\eqref{eq:PolarizabilityRest} widely acknowledged in the literature. The conditions for that include  several relatively close resonances and excitation frequency in the region between them. This provides an interesting parallel with related studies of chiral media~\cite{Tretyakov2003,  Tretyakov2005, Zhang2007, Plum2009, Wang2009, Zhang2009} showing the possibility of strong chirality exceeding the limit $\sqrt{\eps\mu}$.

Even though our conclusion is based on a relatively simple model ignoring losses, we anticipate that the similar violation of restriction Eq.~\eqref{eq:PolarizabilityRest} occurs in realistic Mie-resonant meta-atoms typically featuring multiple spectrally close resonances. Accordingly, the metamaterial composed of such atoms is expected to violate Eq.~\eqref{eq:TellegenRestriction} enabling strong Tellegen response.

Our results provide the theoretical basis for achieving strong Tellegen response in artificial media paving a way to the enhanced nonreciprocal phenomena.



\section*{Acknowledgments}
We acknowledge Maxim Mazanov for valuable discussions. Theoretical models were supported by the Russian Science Foundation, grant No.~23-72-10026. Numerical analysis was supported by Priority 2030 Federal Academic Leadership Program. T.Z.S. and M.A.G. acknowledge partial support from the Foundation for the Advancement of Theoretical Physics and Mathematics “Basis”.

\appendix*

\section{Derivation of the first-order corrections}

Following the approach of Ref.~\cite{Barron_2004}, we calculate the perturbed wave function of the $n$-th stationary state:
\eqref{eq:wf_perturbed}
\begin{multline}
        \ket{n}=\Bigg(\ket{n^{(0)}}+\sum_j \bigg(\alpha_{jn\mu} E_{0\mu}e^{-i\omega t}+ \beta_{jn\mu} E_{0\mu}^\dagger e^{i\omega t} + \\ \xi_{jn\mu} B_{0\mu} e^{-i\omega t}+\zeta_{jn\mu} B_{0\mu}^\dagger e^{i\omega t}\bigg)\ket{j^{(0)}}\Bigg)e^{-i\omega_n t}, \label{eq:wf_perturbed}
\end{multline}
where $\ket{n^{(0)}}$ is the respective eigenstate  of unperturbed Schr{\"o}dinger equation and 
$\alpha_{jn\mu},\beta_{jn\mu},\xi_{jn\mu}$, and $\zeta_{jn\mu}$ are the coefficients to determine. We will refer to the sum in the right-hand side as  $\ket{n^{(1)}}$, as it is the first-order perturbation. To proceed we need to substitute this expansion into Schr{\"o}dinger equation \eqref{eq:Schrodinger}. The term with $\ket{n^{(0)}}$ on the left side cancels out. The rest in the left side is
\begin{multline}
    i\hbar \frac{\partial}{\partial t}\ket{n^{(1)}}e^{-i\omega_n t} = \\
        \hbar \sum_j \Bigg(\alpha_{jn\mu} E_{0\mu}e^{-i\omega t}(\omega_n+\omega)+ \\
        \beta_{jn\mu} E_{0\mu}^\dagger e^{i\omega t}(\omega_n-\omega) + \\
        \xi_{jn\mu} B_{0\mu} e^{-i\omega t}(\omega_n+\omega)+ \\ 
        \zeta_{jn\mu} B_{0\mu}^\dagger e^{i\omega t}(\omega_n-\omega) \ket{j^{(0)}}\Bigg)e^{-i\omega_n t} 
\end{multline}
\begin{multline}
        H \ket{n^{(1)}}=\sum_j \bigg(\alpha_{jn\mu} E_{0\mu}e^{-i\omega t}+ \beta_{jn\mu} E_{0\mu}^\dagger e^{i\omega t} +  \\ \xi_{jn\mu} B_{0\mu} e^{-i\omega t}+\zeta_{jn\mu} B_{0\mu}^\dagger e^{i\omega t}\bigg) \hbar \omega_j \ket{j^{(0)}}e^{-i\omega_n t}
\end{multline}
Introducing $\omega_{jn}=\omega_j-\omega_n$, we sum this up as
\begin{multline}
        -\hbar \sum_j \Bigg(\bigg(\alpha_{jn\mu} E_{0\mu}e^{-i\omega t -i\omega_n t}(\omega_{jn}-\omega)+ \\
        \beta_{jn\mu} E_{0\mu}^\dagger e^{i\omega t -i\omega_n t}(\omega_{jn}+\omega) + \\
        \xi_{jn\mu} B_{0\mu} e^{-i\omega t -i\omega_n t}(\omega_{jn}-\omega)+ \\ 
        \zeta_{jn\mu} B_{0\mu}^\dagger e^{i\omega t -i\omega_n t}(\omega_{jn}+\omega) \bigg) \ket{j^{(0)}}\Bigg)
\end{multline}
In the left hand side $\hat V \ket{n^{(1)}}$ contains only second order terms, which we neglect, leaving only
\begin{multline}
    \hat V \ket{n^{(0)}} e^{-i\omega_n t}= \\ \frac{1}{2}(- d_\beta  E_{0\beta} e^{-i\omega t -i\omega_n t} -  d_\beta  E_{0\beta}^\dagger e^{i\omega t -i\omega_n t} - \\  m_\beta  B_{0\beta} e^{-i\omega t -i\omega_n t} - m_\beta  B_{0\beta}^\dagger e^{i\omega t -i\omega_n t}) \ket{n^{(0)}}
\end{multline}
Multiplying both sides with $\bra{j^{(0)}}$ and equating equal time-exponent parts, we get
\begin{gather}
    \alpha_{j n \mu}=\frac{1}{2\hbar}\frac{\braket{j^{(0)}|d_\mu|n^{(0)}}}{(\omega_{jn}-\omega)} \\
    \beta_{jn\mu}= \frac{1}{2\hbar}\frac{\braket{j^{(0)}|d_\mu|n^{(0)}}}{(\omega_{jn}+\omega)} \\
    \xi_{jn\mu}=\frac{1}{2\hbar}\frac{\braket{j^{(0)}|m_\mu|n^{(0)}}}{(\omega_{jn}-\omega)} \\
    \zeta_{jn\mu}=\frac{1}{2\hbar}\frac{\braket{j^{(0)}|m_\mu|n^{(0)}}}{(\omega_{jn}+\omega)}
\end{gather}
Next we denote for radiation field $E_{0\mu}e^{-i\omega t}\equiv \tilde E_\mu$, $B_{0\mu}e^{-i\omega t}\equiv \tilde B_\mu$, and for the dipole moment matrix elements $\braket{j^{(0)}|d_\mu|n^{(0)}}=d_{jn\mu}$ and $\braket{j^{(0)}|m_\mu|n^{(0)}}=m_{jn\mu}$. Note that according to the definitions, $d_{jn\mu}^*=d_{nj\mu}$, $m_{jn\mu}^*=m_{nj\mu}$ and the true radiation field is $E_\mu=\frac{1}{2}(\tilde E_\mu + \tilde E_\mu^\dagger)$ ( analogously for the magnetic field). Wave function perturbation is
\begin{multline}
   \ket{n^{(1)}}= 
   \frac{1}{2\hbar}\sum_j\Bigg[\skob{2 \omega_{jn} E_\mu+\omega(\tilde E_\mu - \tilde E_\mu^\dagger)}\frac{d_{jn\mu}}{\omega_{jn}^2-\omega^2}+ \\
   \skob{2 \omega_{jn} B_\mu+\omega(\tilde B_\mu - \tilde B_\mu^\dagger)}\frac{m_{jn\mu}}{\omega_{jn}^2-\omega^2}\Bigg] \ket{j^{(0)}}
\end{multline}

Now we are able to calculate expectation value of the perturbed electric dipole moment
\begin{gather}
    \braket{n|d_\nu|n}= \nonumber\\ \braket{n^{(0)}|d_\nu|n^{(0)}}+\braket{n^{(1)}|d_\nu|n^{(0)}}+\braket{n^{(0)}|d_\nu|n^{(1)}}+...=  \nonumber \\
    \braket{n^{(0)}|d_\nu|n^{(0)}}+  \nonumber \\
    \underbrace{\frac{2}{\hbar}\sum_j\frac{\omega_{jn}}{\omega_{jn}^2-\omega^2}\Re{(d_{jn\mu} d_{nj\nu})}}_{\alpha_{\nu\mu}} E_\mu +  \nonumber \\
    \underbrace{\frac{2}{\hbar}\sum_j\frac{1}{\omega_{jn}^2-\omega^2}\Im{(d_{jn\mu} d_{nj\nu})}}_{\alpha_{\nu\mu}'} \dot E_\mu +  \nonumber \\
    \underbrace{\frac{2}{\hbar}\sum_j\frac{\omega_{jn}}{\omega_{jn}^2-\omega^2}\Re{(m_{jn\mu} d_{nj\nu})}}_{\gamma_{\nu\mu}} B_\mu +  \nonumber \\
    \underbrace{\frac{2}{\hbar}\sum_j\frac{1}{\omega_{jn}^2-\omega^2}\Im{(m_{jn\mu} d_{nj\nu})}}_{\gamma_{\nu\mu}'} \dot B_\mu \label{eq:n_d_EV}
\end{gather}
Recalling macroscopic constitutive relation for electric molecule polarization
$d_\nu=\alpha^{ee}_{\nu\mu} E_\mu + \alpha^{em}_{\nu\mu} B_\mu$
we identify $\alpha^{ee}_{\nu\mu}=\alpha_{\nu\mu}+i\alpha_{\nu\mu}'$ and $\alpha^{em}_{\nu\mu}=\gamma_{\nu\mu}+i\gamma_{\nu\mu}'$. A similar procedure for the magnetic moment yields
\begin{gather}
    \braket{n|m_\nu|n}= \nonumber\\ \braket{n^{(0)}|m_\nu|n^{(0)}}+\braket{n^{(1)}|m_\nu|n^{(0)}}+\braket{n^{(0)}|m_\nu|n^{(1)}}+...=  \nonumber \\
    \braket{n^{(0)}|m_\nu|n^{(0)}}+  \nonumber \\
    \underbrace{\frac{2}{\hbar}\sum_j\frac{\omega_{jn}}{\omega_{jn}^2-\omega^2}\Re{(d_{jn\mu} m_{nj\nu})}}_{\gamma_{\nu\mu}} E_\mu +  \nonumber \\
    \underbrace{\frac{2}{\hbar}\sum_j\frac{1}{\omega_{jn}^2-\omega^2}\Im{(d_{jn\mu} m_{nj\nu})}}_{-\gamma_{\nu\mu}'} \dot E_\mu +  \nonumber \\
    \underbrace{\frac{2}{\hbar}\sum_j\frac{\omega_{jn}}{\omega_{jn}^2-\omega^2}\Re{(m_{jn\mu} m_{nj\nu})}}_{\beta_{\nu\mu}} B_\mu +  \nonumber \\
    \underbrace{\frac{2}{\hbar}\sum_j\frac{1}{\omega_{jn}^2-\omega^2}\Im{(m_{jn\mu} m_{nj\nu})}}_{\beta_{\nu\mu}'} \dot B_\mu \label{eq:n_m_EV}
\end{gather}
Macroscopic magnetization is again $m_\nu=\alpha^{me}_{\nu\mu}E_{\mu}+\alpha^{mm}_{\nu\mu} B_{\mu}$ and we thus identify $\alpha^{mm}_{\nu\mu}=\beta_{\nu\mu}+i\beta_{\nu\mu}'$. Moreover, we see that in accordance with energy conservation $\alpha^{me}=\left(\alpha^{em}\right)^\dagger$.

\bibliography{TellegenRestriction}
\end{document}